\documentclass[showpacs,preprintnumbers,
amsmath,amssymb,aps,prd]{revtex4}
\usepackage{}
\usepackage{amsfonts}
\usepackage{amsfonts,graphics,epsfig,subfigure}

\begin{document}

\title{Non-extended phase space thermodynamics of Lovelock AdS black holes in grand canonical ensemble}
\author{Jie-Xiong Mo $^{a,b}$ \footnote{mojiexiong@gmail.com}, Wen-Biao Liu $^a$ \footnote{wbliu@bnu.edu.cn (corresponding
author)}}

 \affiliation{$^a$ Department of Physics, Institute of Theoretical Physics, Beijing Normal University, Beijing, 100875, China\\
 $^b$ Institute of Theoretical Physics, Lingnan Normal University, Zhanjiang, Guangdong, 524048, China\\
  }

\begin{abstract}
   Recently, extended phase space thermodynamics of Lovelock AdS black holes has been of great interest. To provide insight from a different perspective and gain a unified phase transition picture, non-extended phase space thermodynamics of $(n+1)$-dimensional charged topological Lovelock AdS black holes is investigated detailedly in the grand canonical ensemble. Specifically, the specific heat at constant electric potential is calculated and phase transition in the grand canonical ensemble is discussed. To probe the impact of the various parameters, we utilize the control variate method and solve the phase transition condition equation numerically for the case $k=1,-1$. There are two critical points for the case $n=6,k=1$ while there is only one for other cases. For $k=0$, there exists no phase transition point. To figure out the nature of phase transition in the grand canonical ensemble, we carry out an analytic check of the analog form of Ehrenfest equations proposed by Banerjee et al. It is shown that Lovelock AdS black holes in the grand canonical ensemble undergo a second order phase transition. To examine the phase structure in the grand canonical ensemble, we utilize the thermodynamic geometry method and calculate both the Weinhold metric and Ruppeiner metric. It is shown that for both analytic and graphical results that the divergence structure of the Ruppeiner scalar curvature coincides with that of the specific heat. Our research provides one more example that Ruppeiner metric serves as a wonderful tool to probe the phase structures of black holes.
\end{abstract}

  \maketitle

\section{Introduction}
In our recent paper~\cite{Wenbiao3}, $P-V$ criticality of topological AdS black holes in Lovelock-Born-Infeld gravity has been investigated in the extended phase space and some unique phenomena have been found. It was shown that $P-V$ criticality exists not only for the spherical topology but also for $k=-1$. This result is really intriguing that it has attracted further investigation~\cite{Belhaj95}-~\cite{Dolan99}. On the other hand, it would also be interesting to probe this issue in the non-extended phase space to search for some more unique characteristics due to Lovelock gravity. Lovelock gravity~\cite{Lovelock} is a particular higher curvature gravity theory which successfully solves the problem of fourth order field equations and ghost. In Lovelock gravity, the field equation is only second order and the quantization is free of ghosts~\cite{Boulware}. Both the black holes and their thermodynamics in Lovelock gravity~\cite{Dehghani1}-\cite{Amirabi} have attracted considerable attention. Concerning the thermodynamics of Lovelock black holes in the non-extended phase space, some efforts have been made.  Topological black hole solutions in Lovelock-Born-Infeld gravity were proposed in Ref.~\cite{Dehghani1}. Both the thermodynamics of asymptotically AdS rotating black branes with flat horizon and asymptotically flat black holes for $k=1$ were detailedly investigated there. For charged topological AdS black holes, Ref.~\cite{Dehghani1} presented the expression of the temperature. Ref.~\cite{Decheng4,Decheng2} further studied their entropy and specific heat at constant charge. Ref.~\cite{Lala2} studied their specific heat and critical exponents in the canonical ensemble. The above research was carried out in the canonical ensemble, leaving the grand canonical ensemble unexplored. In this paper, we would like to complete the phase transition research of Lovelock charged topological AdS black holes in the grand canonical ensemble.

In traditional thermodynamics, one can utilize Clausius-Clapeyron-Ehrenfest's equations to probe the nature of phase transitions. The Clausius-Clapeyron
equation holds for a first order phase transition while Ehrenfest's equations are satisfied for a second order phase transition. Recently, Banerjee et al. introduced a novel Ehrenfest scheme to investigate phase transitions of black holes in the grand canonical ensemble~\cite{Banerjee1}-\cite{Banerjee6}. We utilized this scheme in the case of charged topological black hole in Ho\v{r}ava-Lifshitz gravity~\cite{Wenbiao2} and also generalized it to the extended phase space~\cite{Wenbiao1}-\cite{Wenbiao4}. Ref.~\cite{Jiliangjing} further generalized it to the full phase space. The original Ehrenfest equations in traditional thermodynamics were utilized in the extended space of black holes in Lovelock-Born-Infeld gravity to study the nature of phase transition at the critical point~\cite{Belhaj95}. However, in this paper, we would like to utilize the analog form of Ehrenfest scheme proposed by Banerjee et al. to investigate the nature of phase transition points of Lovelock AdS black holes in the grand canonical ensemble.

    Different from the traditional thermodynamic method, thermodynamic geometry has served as an alternative way to investigate phase transitions of black holes. The well-known examples are Weinhold geometry~\cite{Weinhold} and Ruppeiner geometry~\cite{Ruppeiner}. Weinhold defined metric structure in the energy representation as $g_{i,j}^{W}=\partial_{i}\partial_{j}M(U,N^a)$. Here, $U$ is the internal energy $U$ while $N^a$ represents the extensive thermodynamic variables. Ruppeiner proposed metric structure as the Hessian of the entropy. Namely, $g_{i,j}^{R}=-\partial_{i}\partial_{j}S(U,N^a)$. Recently, Quevedo et al. \cite{Quevedo2} proposed another thermodynamic geometry method named as geometrothermodynamics (GTD). For its profound physical meaning, Ruppeiner's metric has been applied to investigate various thermodynamic systems including black holes. For a nice review of Ruppeiner geometry, see Ref.~\cite{Ruppeiner2}. For recent papers, see Ref.~\cite{Tharanath}-~\cite{Niu}. However, thermodynamic geometry of Lovelock AdS black holes in the grand canonical ensemble is still absent in literature. In this paper, we would like to explore the Ruppeiner geometry of $(n+1)$-dimensional topological AdS black holes in Lovelock gravity in the grand canonical ensemble .

 In Sec. \ref{Sec2}, the thermodynamics of charged topological AdS black holes in Lovelock-Born-Infeld gravity will be briefly reviewed and the phase transition in the grand canonical ensemble will be investigated in detail. To probe the nature of phase transition in the grand canonical ensemble, an analytic check of the analog form of Ehrenfest equations will be carried out in Sec. \ref{Sec3}. In Sec. \ref{Sec4}, thermodynamic geometry will be studied to examine the phase structure of topological AdS black holes. Concluding remarks will be presented in Sec. \ref {Sec5}.

\section{Phase transition in the grand canonical ensemble}

\label{Sec2}
The action of third order Lovelock-Born-Infeld gravity reads\cite{Dehghani1}
\begin{equation}
I_{G}=\frac{1}{16\pi}\int d^{n+1}x\sqrt{-g}\big(-2\Lambda+\mathcal{L}_1+\alpha_2\mathcal{L}_2+\alpha_3\mathcal{L}_3+L(F)\big),\label{1}
\end{equation}%
where
\begin{eqnarray}
\mathcal{L}_1&=&R,\label{2}
\\
\mathcal{L}_2&=&R_{\mu\nu\gamma\delta}R^{\mu\nu\gamma\delta}-4R_{\mu\nu}R^{\mu\nu}+R^2, \label{3}
\\
\mathcal{L}_3&=&2R^{\mu\nu\sigma\kappa}R_{\sigma\kappa\rho\tau}R^{\rho\tau}_{\;\;\;\;\mu\nu}+8R^{\mu\nu}_{\;\;\;\;\sigma\rho}R^{\sigma\kappa}_{\;\;\;\;\nu\tau}R^{\rho\tau}_{\;\;\;\;\mu\kappa}+24R^{\mu\nu\sigma\kappa}R_{\sigma\kappa\nu\rho}R^{\rho}_{\;\;\mu} \nonumber
\\
&\,&+3RR^{\mu\nu\sigma\kappa}R_{\sigma\kappa\mu\nu}+24R^{\mu\nu\sigma\kappa}R_{\sigma\mu}R_{\kappa\nu}+16R^{\mu\nu}R_{\nu\sigma}R^{\sigma}_{\;\;\mu}-12RR^{\mu\nu}R_{\mu\nu}+R^3,\label{4}
\\
L(F)&=&4\beta^2\left(1-\sqrt{1+\frac{F^2}{2\beta^2}}\right). \label{5}
\end{eqnarray}%
$\beta$, $\alpha_2$ and $\alpha_3$ are Born-Infeld parameter, the second and third order Lovelock coefficients respectively. $L(F)$ denotes the Born-Infeld Lagrangian with $F_{\mu\nu}=\partial_\mu A_\nu-\partial_\nu A_\mu$, where $A_\mu$ is electromagnetic vector. The $(n+1)$-dimensional static solution was derived in Ref.~\cite{Dehghani1} as
\begin{equation}
ds^2=-f(r)dt^2+\frac{dr^2}{f(r)}+r^2d\Omega^2, \label{6}
\end{equation}%
where
\begin{eqnarray}
f(r)&=&k+\frac{r^2}{\alpha}(1-g(r)^{1/3}),\label{7}\\
g(r)&=&1+\frac{3\alpha m}{r^n}-\frac{12\alpha \beta^2}{n(n-1)}\Big[1-\sqrt{1+\eta}-\frac{\Lambda}{2\beta^2}+\frac{(n-1)\eta}{(n-2)}\digamma(\eta)\Big].\label{8}
\end{eqnarray}%
$k$ and $m$ are parameters related to the curvature of hypersurface and the mass respectively. $d\Omega^2$ denotes the line element of $(n-1)$-dimensional hypersurface with constant curvature $(n-1)(n-2)k$ and $\digamma(\eta)$ denotes the hypergeometric function as follow
\begin{equation}
\digamma(\eta)=\,_2F_1\Big(\Big[\frac{1}{2},\frac{n-2}{2n-2}\Big],\Big[\frac{3n-4}{2n-2}\Big],-\eta\Big), \label{9}
\end{equation}%
where
\begin{equation}
\eta=\frac{(n-1)(n-2)q^2}{2\beta^2r^{2n-2}}. \label{10}
\end{equation}%
Note that the above solution was derived for the special case that
\begin{eqnarray}
\alpha_2&=&\frac{\alpha}{(n-2)(n-3)}, \label{11}
\\
\alpha_3&=&\frac{\alpha^2}{72{n-2\choose 4}}.\label{12}
\end{eqnarray}%

When $\beta\rightarrow\infty$, the Born-Infeld Lagrangian reduces to the Maxwell form and the solutions become Lovelock AdS black holes. To concentrate on the effects of the third order Lovelock gravity, we will mainly consider Lovelock AdS black holes in this paper.

When $\beta\rightarrow\infty$, one can obtain
\begin{equation}
g(r)\rightarrow1+\frac{3\alpha m}{r^n}+\frac{6\alpha \Lambda}{n(n-1)}-\frac{3\alpha q^2}{r^{2n-2}}.\label{13}
\end{equation}%
The horizon radius $r_+$ can be derived from the largest root of the equation $f(r)=0$. One can express $m$ in the function of $r_+$ as
\begin{equation}
m=\frac{3n(n-1)q^2r_+^8+r_+^{2n}\left[kn(n-1)(3r_+^4+3kr_+^2\alpha+k^2\alpha^2)-6r_+^6\Lambda\right]}{3n(n-1)r_+^{n+6}}.\label{14}
\end{equation}%
Then the mass of $(n+1)$-dimensional topological AdS black holes can be derived as
\begin{equation}
M=\frac{(n-1)\Sigma_k}{16\pi}m=\frac{\Sigma_k}{48n\pi r_+^{n+6}}\left\{3n(n-1)q^2r_+^8+r_+^{2n}\left[kn(n-1)(3r_+^4+3kr_+^2\alpha+k^2\alpha^2)-6r_+^6\Lambda\right]\right\},\label{15}
\end{equation}%
where $\Sigma_k$ denotes the volume of the $(n-1)$-dimensional hypersurface.
The Hawking temperature has been derived in Ref.~\cite{Dehghani1} as
\begin{equation}
T=\frac{(n-1)k[3(n-2)r_+^4+3(n-4)k\alpha r_+^2+(n-6)k^2\alpha^2]+12r_+^6\beta^2(1-\sqrt{1+\eta_+}\,)-6\Lambda r_+^6}{12\pi(n-1)r_+(r_+^2+k\alpha)^2}.\label{16}
\end{equation}%
Taking the limit $\beta\rightarrow\infty$, Eq. (\ref{16}) reduces to
\begin{equation}
T=\frac{(n-1)k[3(n-2)r_+^4+3(n-4)k\alpha r_+^2+(n-6)k^2\alpha^2]-6\Lambda r_+^6-3(n-2)(n-1)q^2r_+^{8-2n}}{12\pi(n-1)r_+(r_+^2+k\alpha)^2}.\label{17}
\end{equation}%
In the non-extended phase space, the first law of thermodynamics reads
\begin{equation}
dM=TdS+\Phi dQ.\label{18}
\end{equation}%
So the entropy can be derived as
\begin{equation}
S=\int^{r_+}_{0}\frac{1}{T}\left(\frac{\partial M}{\partial r_+}\right)dr=\frac{\Sigma_k(n-1)r_+^{n-5}}{4}\left(\frac{r_+^4}{n-1}+\frac{2kr_+^2\alpha}{n-3}+\frac{k^2\alpha^2}{n-5}\right).\label{19}
\end{equation}%
 The above result is derived for $n>5$ while the integration is divergent for $n\leqslant5$.
The charge $Q$ is related to the parameter $q$ by
\begin{equation}
Q=\frac{\Sigma_k}{4\pi}\sqrt{\frac{(n-1)(n-2)}{2}}q.\label{20}
\end{equation}%
Then the expression of the mass can be reorganized as
\begin{equation}
M=\frac{96n\pi ^2 Q^2r_+^8+r_+^{2n}(n-2)\left[kn(n-1)(3r_+^4+3kr_+^2\alpha+k^2\alpha^2)-6r_+^6\Lambda\right]\Sigma_k^2}{48n(n-2)\pi r_+^{n+6}\Sigma_k}.\label{21}
\end{equation}%
Utilizing Eqs.~(\ref{18}) and (\ref{21}), the electric potential can be calculated as
\begin{equation}
\Phi=\Big(\frac{\partial M}{\partial Q}\Big)_{S}=\frac{4\pi Q}{(n-2)r_+^{n-2}\Sigma_k}.\label{22}
\end{equation}%

To study the phase transition in the grand canonical ensemble, it is more convenient to express the mass into the function of the electric potential $\Phi$ as follows
\begin{eqnarray}
M&=&\frac{6n(n-2)^2 \Phi^2r_+^{2n+4}\Sigma_k^3+r_+^{2n}(n-2)\Sigma_k\left[kn(n-1)(3r_+^4+3kr_+^2\alpha+k^2\alpha^2)-6r_+^6\Lambda\right]}{48n(n-2)\pi r_+^{n+6}},\label{23}\\
T&=&\frac{(n-1)k[3(n-2)r_+^4+3(n-4)k\alpha r_+^2+(n-6)k^2\alpha^2]-6\Lambda r_+^6-6(n-2)^2\Phi^2r_+^4}{12\pi(n-1)r_+(r_+^2+k\alpha)^2}.\label{24}
\end{eqnarray}%

\begin{figure*}
\centerline{\subfigure[]{\label{2a}
\includegraphics[width=8cm,height=6cm]{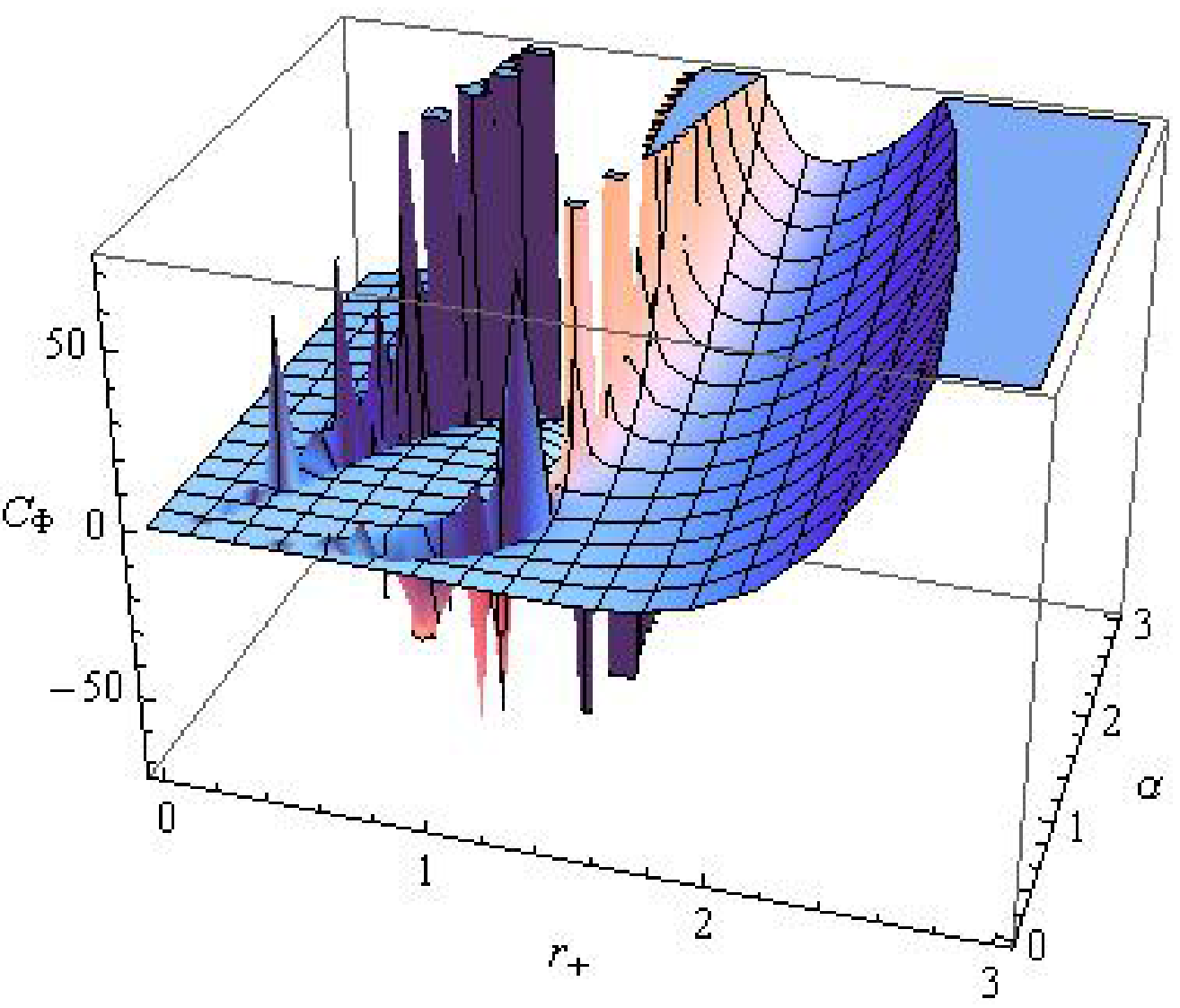}}
\subfigure[]{\label{2b}
\includegraphics[width=8cm,height=6cm]{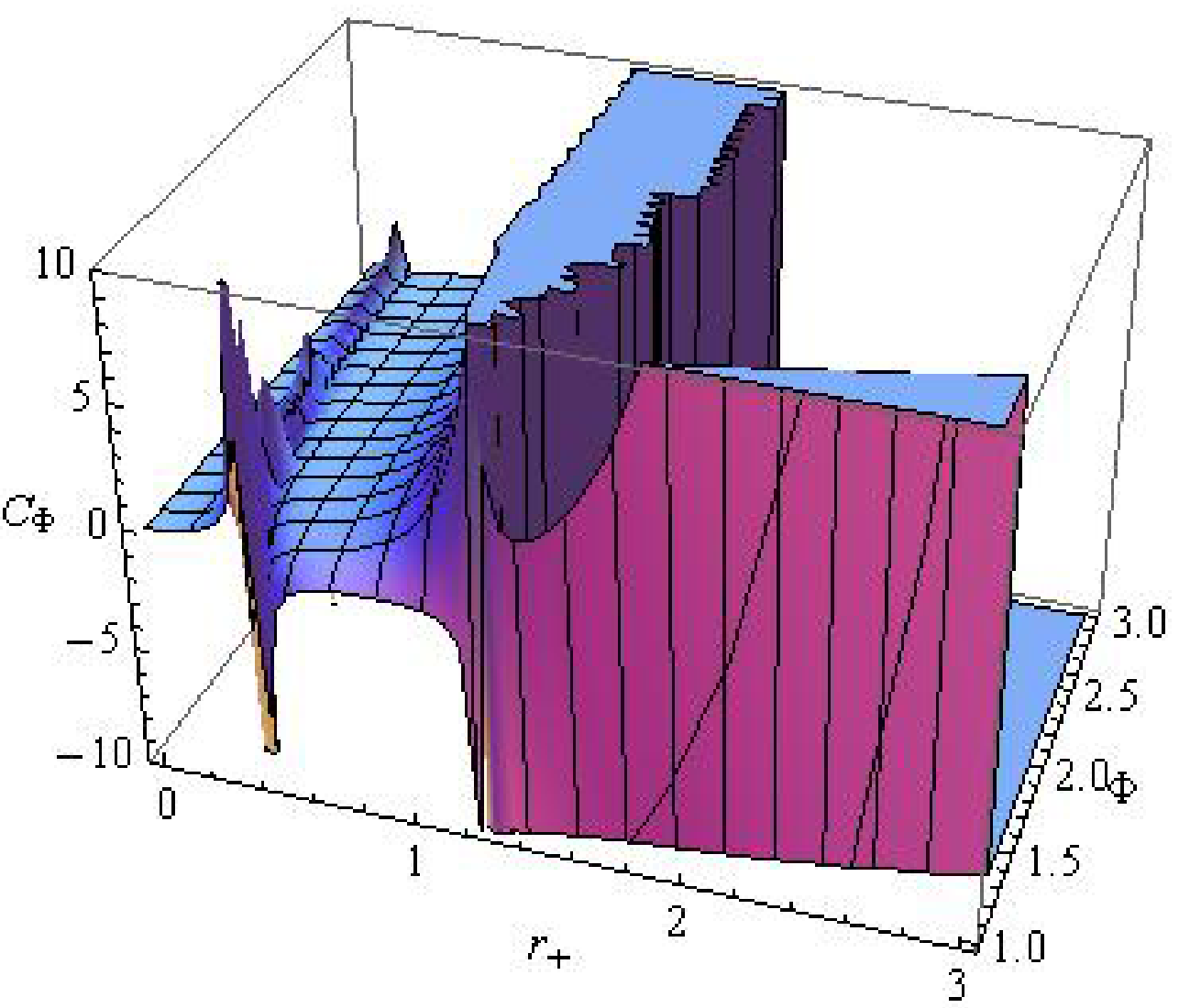}}}
\centerline{\subfigure[]{\label{2c}
\includegraphics[width=8cm,height=6cm]{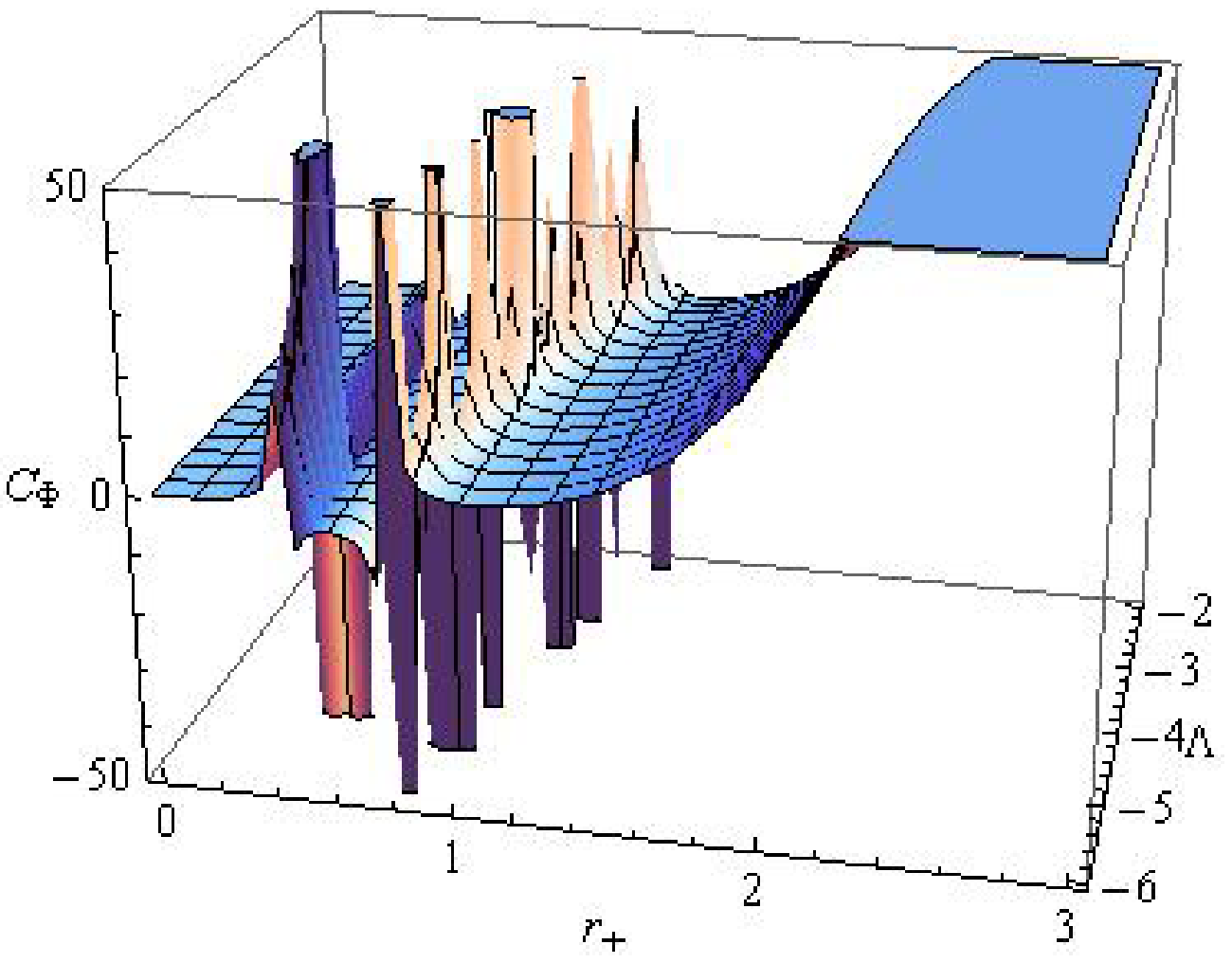}}}
 \caption{(a) $C_{\Phi}$ vs. $r_+$ for $k=1, n=6,\Phi=1,\Lambda=-2$ (b) $T$ vs. $r_+$ for $k=1, n=6,\alpha=1,\Lambda=-2$ (c) $C_{\Phi}$ vs. $r_+$ for $k=1, n=6,\alpha=1,\Phi=1$} \label{fg2}
\end{figure*}

\begin{figure*}
\centerline{\subfigure[]{\label{3a}
\includegraphics[width=8cm,height=6cm]{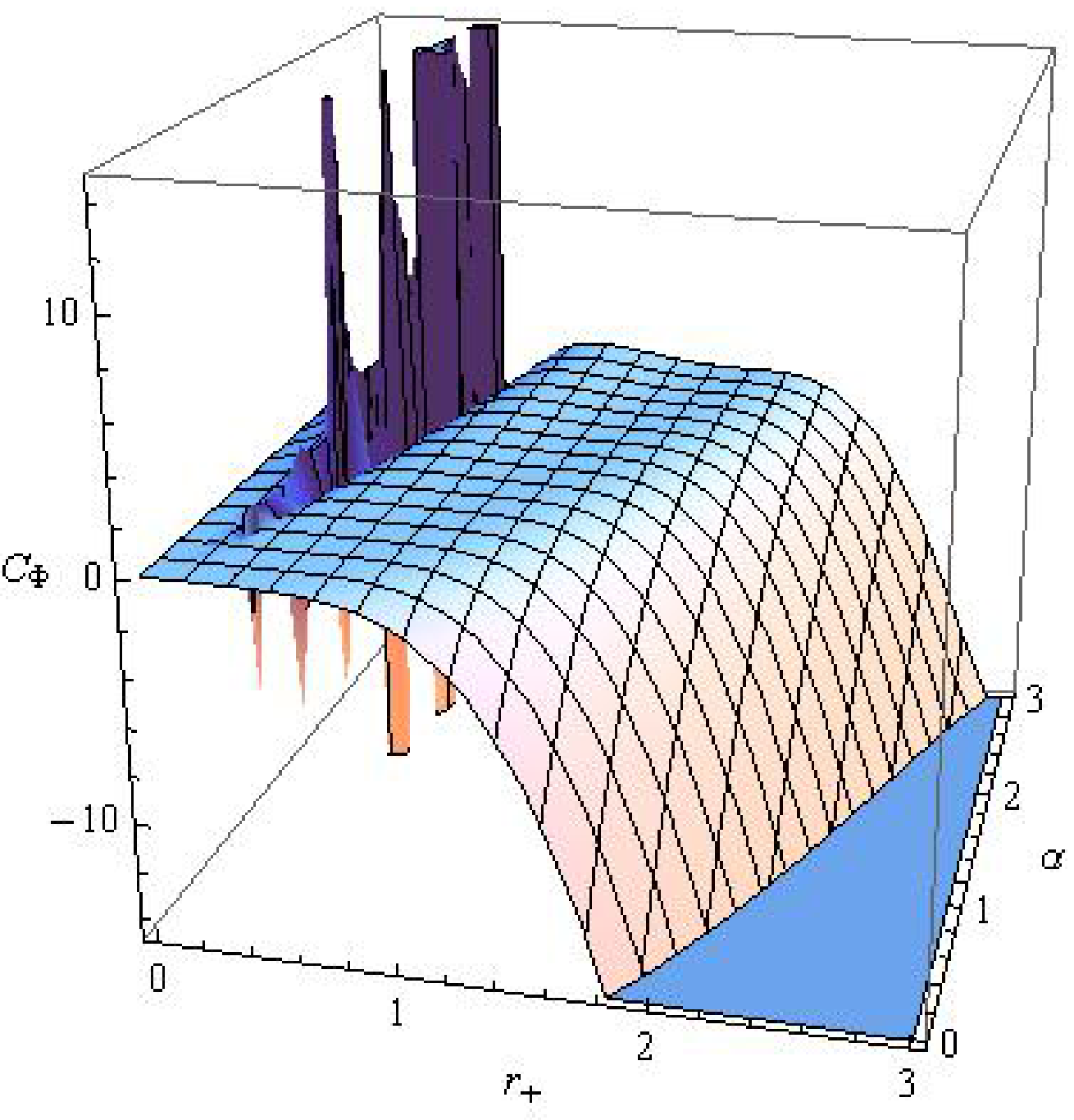}}
\subfigure[]{\label{3b}
\includegraphics[width=8cm,height=6cm]{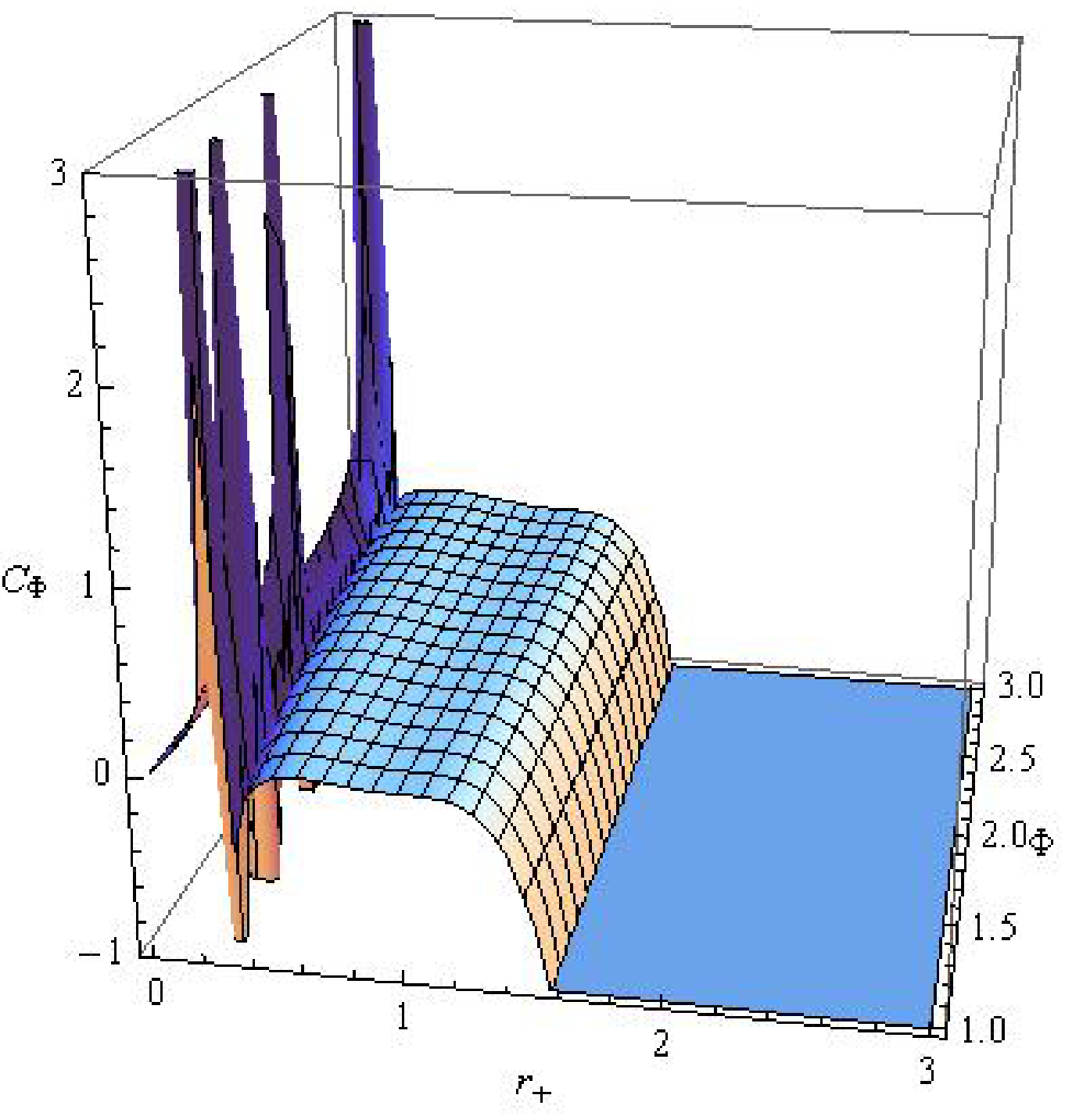}}}
\centerline{\subfigure[]{\label{3c}
\includegraphics[width=8cm,height=6cm]{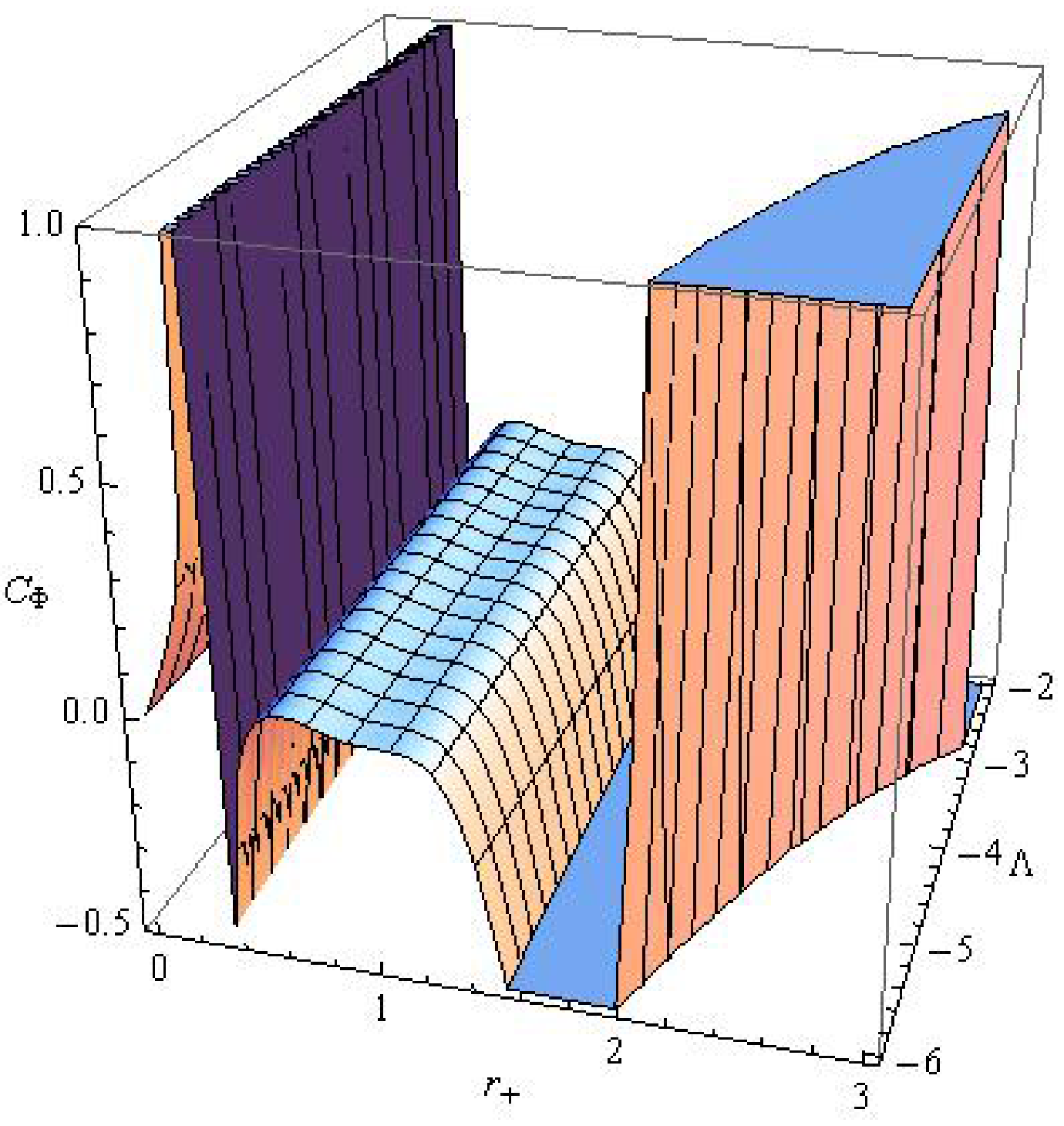}}}
 \caption{(a) $C_{\Phi}$ vs. $r_+$ for $k=-1, n=6,\Phi=1,\Lambda=-2$ (b) $T$ vs. $r_+$ for $k=-1, n=6,\alpha=1,\Lambda=-2$ (c) $C_{\Phi}$ vs. $r_+$ for $k=-1, n=6,\alpha=1,\Phi=1$} \label{fg3}
\end{figure*}

The specific heat at constant electric potential can be obtained as
\begin{equation}
C_{\Phi}=T\Big(\frac{\partial S}{\partial T}\Big)_{\Phi}=\frac{A(r_+,\Phi)}{B(r_+,\Phi)},\label{25}
\end{equation}%
where
\begin{eqnarray}
A(r_+,\Phi)&=&(n-1)r_+^{n-5}(r_+^2+k\alpha)^3\Sigma_k
\nonumber
\\
&\,&\times\left[3k(n-2)(n-1)r_+^4+3k^2(n-4)(n-1)r_+^2\alpha+k^3(n-6)(n-1)\alpha^2-6r_+^6\Lambda-6(n-2)^2r_+^4\Phi^2\right],\label{26}\\
B(r_+,\Phi)&=&-24\Lambda r_+^8-12\left[k(2+n^2-3n+10\alpha\Lambda)-2(n-2)^2\Phi^2\right]r_+^6+72k\alpha r_+^4\left[k(n-1)-(n-2)^2\Phi^2\right]
\nonumber
\\
&\,&-8k^3(n-9)(n-1)\alpha^2r_+^2-4k^4(n-6)(n-1)\alpha^3.\label{27}
\end{eqnarray}%
One can easily draw the conclusion that the specific heat at constant electric potential may diverge when
\begin{equation}
B(r_+,\Phi)=0,\label{28}
\end{equation}%
implying the existence of phase transition.
\begin{figure*}
\centerline{\subfigure[]{\label{1a}
\includegraphics[width=8cm,height=6cm]{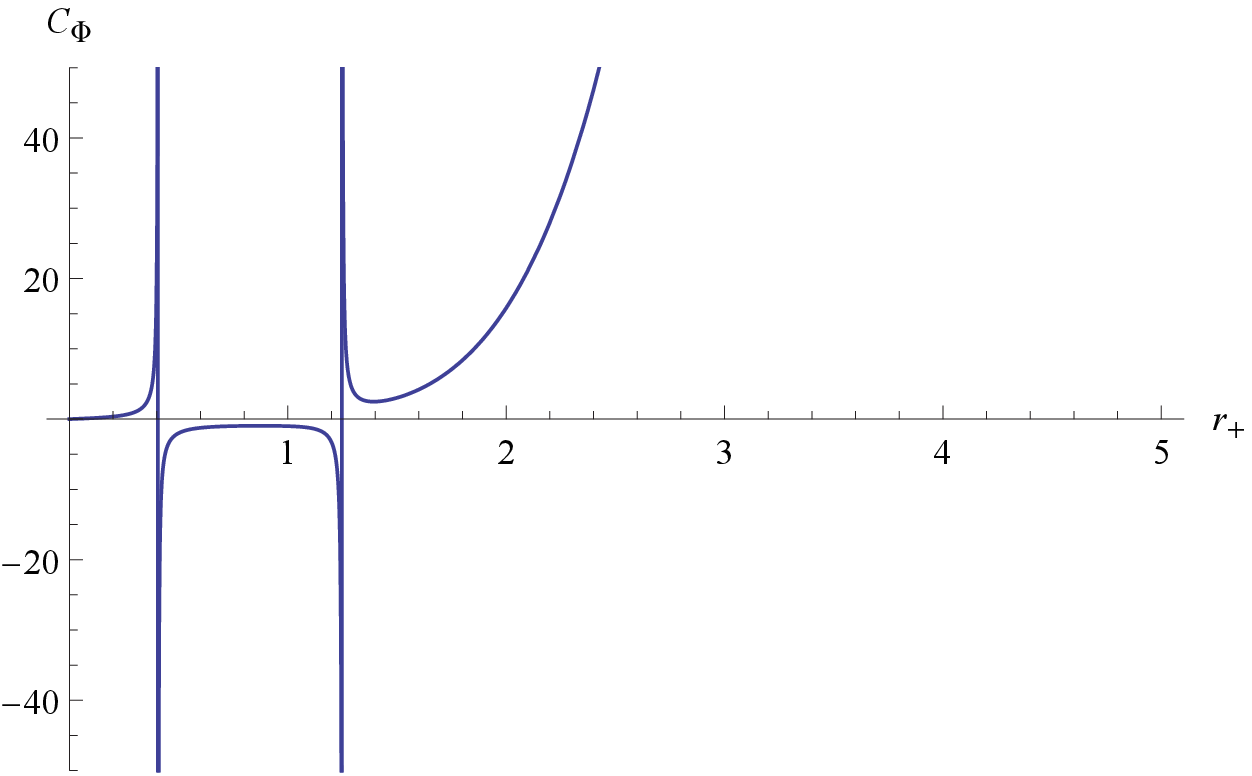}}
\subfigure[]{\label{1b}
\includegraphics[width=8cm,height=6cm]{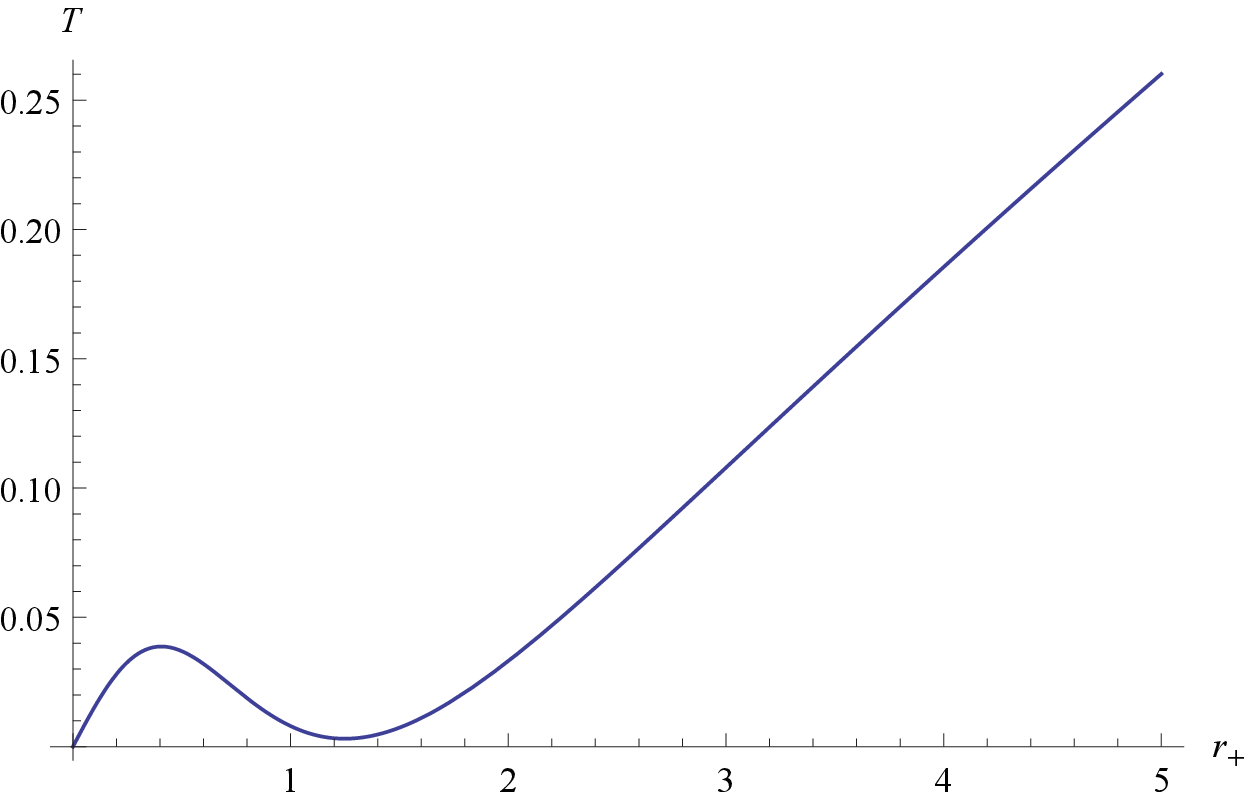}}}
 \caption{(a) $C_{\Phi}$ vs. $r_+$ for $k=1, n=6,\alpha=1,\Phi=1,\Lambda=-2$ (b) $T$ vs. $r_+$ for $k=1, n=6,\alpha=1,\Phi=1,\Lambda=-2$} \label{fg1}
\end{figure*}
The above equation can be solved numerically and the results for $k=1,-1$ are presented in Table \ref{tb1}-\ref{tb2} respectively, where the impact of the various parameters are studied thoroughly via control variate method. It is quite interesting to note that there are two critical points for the case $n=6,k=1$ while there is only one for other cases. And the distance between the two phase transition points becomes larger with the increasing of $\Lambda$ and $\Phi$ while it first becomes larger then becomes smaller with the increasing of $\alpha$. The case $n=6,k=1$ is shown graphically in Fig.\ref{fg1}. Both the behaviors of the specific heat and Hawking temperature are depicted. It is easy to find that the two phase transition points where the specific heat diverges are physical for the Hawking temperature is positive. The black holes can be divided into three phases. Namely, small stable ($C_\Phi>0$) black hole, medium unstable ($C_\Phi<0$) black hole and large stable ($C_\Phi>0$) black hole. For a more comprehensive picture, we also plot the three-dimensional figure for the case $n=6,k=1$ in Fig.\ref{fg2} and for the case $n=6,k=-1$ in Fig.\ref{fg3}.

The case $k=0$ is quite simple. When $k=0$, Eq.~(\ref{27}) can be simplified as
\begin{equation}
B(r_+,\Phi)=-24\Lambda r_+^8+24(n-2)^2\Phi^2>0.\label{29}
\end{equation}%
So there exists no phase transition for $k=0$.
\begin{table}[!h]
\tabcolsep 0pt
\caption{The location of phase transition points for $k=1$}
\vspace*{-12pt}
\begin{center}
\def\temptablewidth{0.5\textwidth}
{\rule{\temptablewidth}{1pt}}
\begin{tabular*}{\temptablewidth}{@{\extracolsep{\fill}}cccccc}
$n$ & $\alpha$ & $\Phi$ &$\Lambda$ &$r_{+1}$ &$r_{+2}$ \\   \hline
     6  & 1 &1 &       -2& 0.406& 1.248\\
       6     &1  & 1&      -4 & 0.420& 0.964 \\
       6     & 1  &1       & -6& 0.440  & 0.794 \\
           6     & 0.5  &1&        -2& 0.283 & 1.076 \\
              6     & 2  &1&        -2& 0.595 & 1.363  \\
                6     & 1  &2&        -2& 0.169 & 1.592 \\
                 6     & 1  &3&        -2& 0.110 & 1.666  \\
                 7   & 1 &1&        -2& 1.503  & -\\
                   8     & 1  &1&        -2& 1.680  & - \\
                    9     & 1  &1&       -2& 1.812 & -  \\
       \end{tabular*}
       {\rule{\temptablewidth}{1pt}}
       \end{center}
       \label{tb1}
       \end{table}

\begin{table}[!h]
\tabcolsep 0pt
\caption{The location of phase transition points for $k=-1$}
\vspace*{-12pt}
\begin{center}
\def\temptablewidth{0.5\textwidth}
{\rule{\temptablewidth}{1pt}}
\begin{tabular*}{\temptablewidth}{@{\extracolsep{\fill}}ccccc}
$n$ & $\alpha$ & $\Phi$ &$\Lambda$ &$r_{+}$  \\   \hline
     6  & 1 &1 &       -2& 0.279\\
       6     &1  & 1&      -4 & 0.281 \\
       6     & 1  &1       & -6& 0.282  \\
           6     & 0.5  &1&        -2& 0.197 \\
              6     & 2  &1&        -2& 0.397   \\
                6     & 1  &2&        -2& 0.155  \\
                 6     & 1  &3&        -2& 0.106   \\
                 7   & 1 &1&        -2& 0.353  \\
                   8     & 1  &1&        -2& 0.374   \\
                    9     & 1  &1&       -2& 0.386   \\
       \end{tabular*}
       {\rule{\temptablewidth}{1pt}}
       \end{center}
       \label{tb2}
       \end{table}
\section{The nature of phase transition in the grand canonical ensemble}
\label{Sec3}

\begin{figure*}
\centerline{\subfigure[]{\label{4a}
\includegraphics[width=8cm,height=6cm]{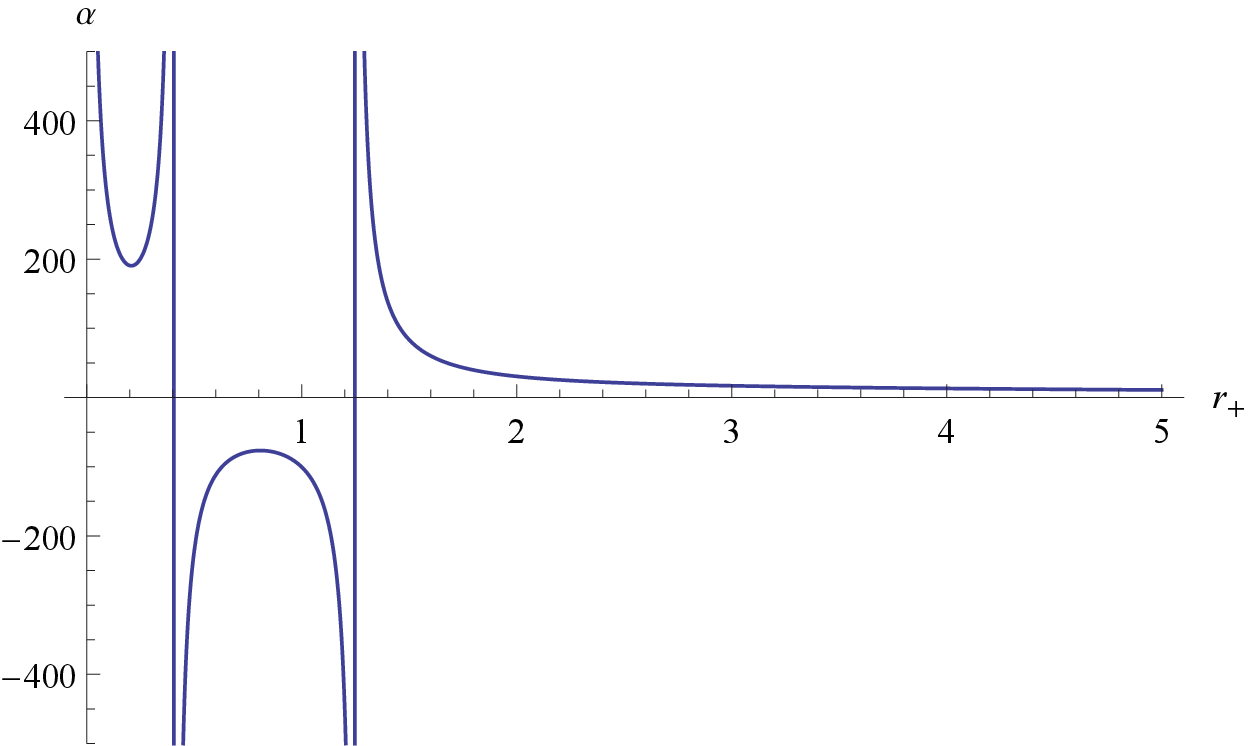}}
\subfigure[]{\label{4b}
\includegraphics[width=8cm,height=6cm]{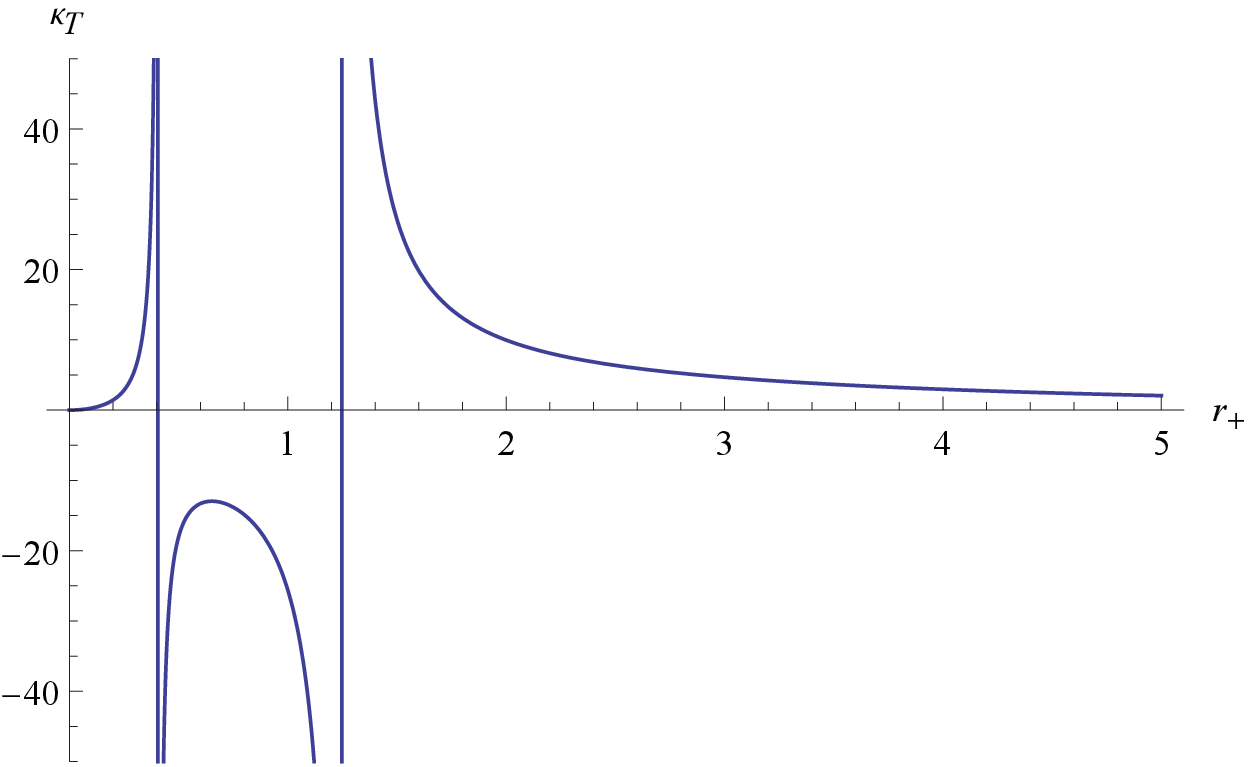}}}
 \caption{(a) $\alpha$ vs. $r_+$ for $k=1, n=6,\alpha=1,\Phi=1,\Lambda=-2$ (b) $\kappa_T$ vs. $r_+$ for $k=1, n=6,\alpha=1,\Phi=1,\Lambda=-2$} \label{fg4}
\end{figure*}

In the extended space, it is convenient to utilize the classical Ehrenfest equations to study the nature of phase transition at the critical point. However, here, in the non-extended phase space, we would like to introduce the novel analog form of Ehrenfest equations proposed by Banerjee et al.~\cite{Banerjee1} as follow
\begin{eqnarray}
-\left(\frac{d \Phi}{d T}\right)_S&=&\frac{C_{\Phi_2}-C_{\Phi_1}}{TQ(\alpha_2-\alpha_1)}=\frac{\Delta
C_\Phi}{TQ\Delta \alpha},\label{30}\\
-\left(\frac{d \Phi}{d T}\right)_Q&=&\frac{\alpha_2-\alpha_1}{\kappa_{T_2}-\kappa_{T_1}}=\frac{\Delta
\alpha}{\Delta\kappa_T},\label{31}
\end{eqnarray}%
where $\alpha=\frac{1}{Q}(\frac{\partial Q}{\partial T})_\Phi$, $\kappa_T=\frac{1}{Q}(\frac{\partial Q}{\partial \Phi})_T$ are
the analog of volume expansion coefficient and isothermal compressibility respectively. Their explicit forms can be calculated as follows
\begin{eqnarray}
\alpha&=&\frac{48(n-1)(n-2)\pi r_+(r_+^2+k\alpha)^3}{B(r_+,\Phi)},\label{32}\\
\kappa_T&=&\frac{48(n-2)^3r_+^6(r_+^2+k\alpha)\Phi}{B(r_+,\Phi)}.\label{33}
\end{eqnarray}%
$\alpha,\kappa_T$ may also diverge at the phase transition point because they share
the same factor as $C_\Phi$ in their denominators. It can be clearly seen in Fig.~\ref{fg4}.

From the definitions of $\alpha$ and $C_\Phi$, one can obtain
\begin{equation}
Q\alpha=(\frac{\partial Q}{\partial T})_\Phi=(\frac{\partial
Q}{\partial S})_\Phi(\frac{\partial S}{\partial
T})_\Phi=(\frac{\partial Q}{\partial S})_\Phi(\frac{C_\Phi
}{T}).\label{34}
\end{equation}%
So the R.H.S of Eq.(\ref{30}) can be derived as
\begin{equation}
\frac{\Delta C_\Phi}{TQ\Delta \alpha}=(\frac{\partial S}{\partial
Q})_\Phi=\frac{(n-1)\pi(r_+^2+k\alpha)^2}{(n-2)^2r_+^3\Phi}.\label{35}
\end{equation}%
On the other hand, the L.H.S of Eq.(\ref{30}) can be derived as
\begin{equation}
-(\frac{\partial \Phi}{\partial T})_S=\frac{(n-1)\pi(r_+^2+k\alpha)^2}{(n-2)^2r_+^3\Phi}.\label{36}
\end{equation}%
    So the first equation of Erhenfest equations has been proved to be valid.

    The L.H.S of Eq.(\ref{31}) can be obtained as
\begin{equation}
-\left(\frac{\partial \Phi}{\partial T}\right)_Q=\frac{-1}{\left(\frac{\partial T}{\partial \Phi}\right)_Q}=\frac{-1}{\left(\frac{\partial T}{\partial \Phi}\right)_{r_+}+\left(\frac{\partial T}{\partial r_+}\right)_\Phi \left(\frac{\partial r_+}{\partial \Phi}\right)_Q}=\frac{-1}{\left(\frac{\partial T}{\partial \Phi}\right)_{r_+}}=\frac{(n-1)\pi(r_+^2+k\alpha)^2}{(n-2)^2r_+^3\Phi}.\label{37}
\end{equation}%
Note that we have utilized the phase transition condition $\left(\frac{\partial T}{\partial r_+}\right)_\Phi=0$. From the thermodynamic identity \cite{Banerjee5}
\begin{equation}
(\frac{\partial Q}{\partial \Phi})_T(\frac{\partial \Phi}{\partial
T})_Q(\frac{\partial T}{\partial Q})_\Phi=-1,\label{38}
\end{equation}%
we can derive that
\begin{equation}
Q\kappa_T=(\frac{\partial Q}{\partial \Phi})_T=-(\frac{\partial
T}{\partial \Phi})_Q(\frac{\partial Q}{\partial
T})_\Phi=-(\frac{\partial T}{\partial \Phi})_QQ\alpha.\label{39}
\end{equation}%
Noting that in the above derivation, we have also utilized both the definitions of $\kappa_T$ and $\alpha$. We can obtain
\begin{equation}
\frac{\Delta \alpha}{\Delta \kappa_T}=-(\frac{\partial
\Phi}{\partial T})_Q=\frac{(n-1)\pi(r_+^2+k\alpha)^2}{(n-2)^2r_+^3\Phi}.\label{40}
\end{equation}%
               From Eqs.(\ref{37}) and (\ref{40}), one can easily draw the conclusion that the second equation of Ehrenfest equations also holds. The Prigogine-Defay(PD)
                ratio can be calculated as
\begin{equation}
\Pi=\frac{\Delta C_\Phi \Delta \kappa_T}{T_cQ(\Delta
\alpha)^2}=1.\label{41}
\end{equation}%
                Eq.(\ref{41}) and the validity of Ehrenfest equations show that Lovelock AdS black holes in grand canonical ensemble undergo second order phase transition.

\section{Thermodynamic geometry of Lovelock AdS black holes}
\label{Sec4}
Weinhold's metric~\cite{Weinhold} and Ruppeiner's metric~\cite{Ruppeiner} are defined respectively as
\begin{eqnarray}
g_{ij}^{W}&=&\frac{\partial ^2 U(x^k)}{\partial x^i
\partial x^j},\label{42}
\\
g_{ij}^{R}&=&-\frac{\partial ^2 S(x^k)}{\partial x^i
\partial x^j}.\label{43}
\end{eqnarray}
And they are conformally connected to each other through the map \cite{Janyszek}
\begin{equation}
dS^2_R=\frac{dS^2_W}{T}.\label{44}
\end{equation}

\begin{figure}
\includegraphics[width=8cm,height=6cm]{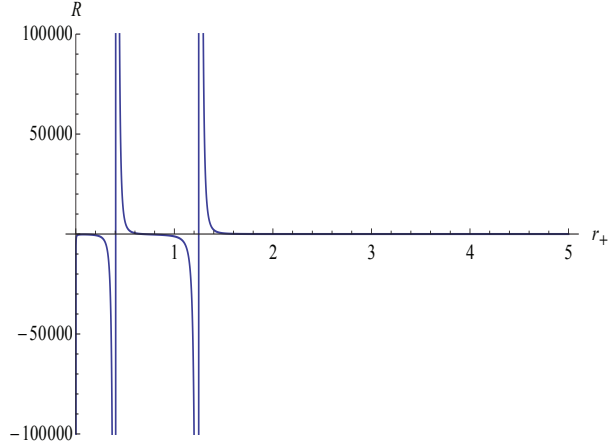}
 \caption{Ruppeiner scalar curvature $R$ vs. $r_+$ for $k=1, n=6,\alpha=1,\Phi=1,\Lambda=-2$}
\label{fg5}
\end{figure}

Utilizing Eqs.(\ref{19}) and (\ref{21}), one can obtain the components of Weinhold's metric as

\begin{eqnarray}
g_{SS}^{W}&=&\frac{D(r_+,Q)}{3(n-1)^2\pi r_+^{3n-4}(r_+^2+k\alpha)^5\Sigma_k^3},\label{45}
\\
g_{QQ}^{W}&=&\frac{4\pi}{(n-2)r_+^{n-2}\Sigma_k},\label{46}
\\
g_{SQ}^{W}&=&g_{QS}^{W}=\frac{-16\pi Q}{(n-1)r_+^{2n-7}(r_+^2+k\alpha)^2\Sigma_k^2},\label{47}
\end{eqnarray}%
where
\begin{eqnarray}
D(r_+,Q)&=&96\pi^2Q^2r_+^8\left[(2n-3)r_+^2+k(2n-7)\alpha\right]-r_+^{2n}
\nonumber
\\
&\,&\times\left\{k(n-1)\left[3(n-2)r_+^6-18kr_+^4\alpha+2k^2(n-9)r_+^2\alpha^2+k^3(n-6)\alpha^3\right]+6r_+^6\Lambda(r_+^2+5k\alpha)\right\}\Sigma_k^2.\label{48}
\end{eqnarray}%
Utilizing Eqs.(\ref{17}), (\ref{44}), (\ref{45}), (\ref{46}) and
(\ref{47}), the components of Ruppeiner's metric can be derived as

\begin{eqnarray}
g_{11}^{R}&=&\frac{4r_+^{5-3n}D(r_+,Q)}{(n-1)(r_+^2+k\alpha)^3\left\{k(n-1)\left[3(n-2)r_+^4+3k(n-4)r_+^2\alpha+k^2(n-6)\alpha^2\right]\Sigma_k^3-6r_+^6\Lambda\Sigma_k^3-96\pi^2Q^2r_+^{8-2n}\Sigma_k\right\}},
\nonumber
\\
\label{49}
\\
g_{22}^{R}&=&\frac{48(n-1)\pi^2r_+^{n+3}(r_+^2+k\alpha)^2\Sigma_k}{(n-2)\left\{-96\pi^2Q^2r_+^8+r_+^{2n}(n-1)k\left[3(n-2)r_+^4+3k(n-4)r_+^2\alpha+k^2(n-6)\alpha^2\right]\Sigma_k^2-6r_+^{2n+6}\Lambda\Sigma_k^2\right\}},\label{50}
\\
g_{12}^{R}&=&g_{21}^{R}=\frac{-192\pi^2 Qr_+^{8-2n}}{k(n-1)\left[3(n-2)r_+^4+3k(n-4)r_+^2\alpha+k^2(n-6)\alpha^2\right]\Sigma_k^2-6r_+^6\Lambda\Sigma_k^2-96\pi^2Q^2r_+^{8-2n}}.\label{51}
\end{eqnarray}%
Utilizing Eqs.(\ref{49})-(\ref{51}), we can obtain Ruppeiner scalar
curvature as
\begin{equation}
R=\frac{E(r_+,\Phi)}{F(r_+,\Phi)},\label{53}
\end{equation}%
where
\begin{eqnarray}
F(r_+,\Phi)&=&-(n-1)r_+^n(r_+^2+k\alpha)^3\Sigma_k\times\{-6\Lambda r_+^8-3r_+^6\left[k(2+n^2-3n+10\alpha\Lambda)-2(n-2)^2\Phi^2\right]
\nonumber
\\
&\,&+18k\alpha r_+^4\left[k(n-1)-(n-2)^2\Phi^2\right]-2k^3(n-9)(n-1)\alpha^2r_+^2-k^4(n-6)(n-1)\alpha^3\}^3
\nonumber
\\
&\,&\times \left[3k(n-2)(n-1)r_+^4+3k^2(n-4)(n-1)r_+^2\alpha+k^3(n-6)(n-1)\alpha^2-6r_+^6\Lambda-6(n-2)^2r_+^4\Phi^2 \right],\label{54}
\end{eqnarray}%
and $E(r_+,Q)$ is too lengthy to be displayed here. The above result has been rewritten in the function of $\Phi$ so that we can compare it with the specific heat. It is not difficult to observe from Eq.(\ref{53}) that in the denominator of Ruppeiner scalar curvature, the fifth factor is exactly one quarter of the denominator of the specific heat while the last factor coincides with the numerator of the Hawking temperature. In other words, the Ruppeiner scalar curvature may diverge exactly where the specific heat diverges. It also reveals the extremal black hole condition that the Hawking temperature is zero. For an intuitive understanding, one can observe the behavior of Ruppeiner scalar curvature $R$ in Fig.\ref{fg5}. Comparing Fig.\ref{fg5} with Fig.\ref{1a}, one can find that the divergence structures of both the Ruppeiner scalar curvature and the specific heat are exactly the same. And the Ruppeiner metric does provide a wonderful tool for one to probe the phase structures of black holes.

Among thermodynamic geometry theories, Ruppeiner geometry has been proved to be outstanding for its profound physical meaning. As argued in Ref. \cite{Ruppeiner2}, Ruppeiner scalar curvature $R$ results from the thermodynamic information metric giving thermodynamic fluctuations and may be interpreted physically as the measurement of the correlation between fluctuating Planck length pixels of event horizon. In the region with positive $R$, repulsive interactions (Fermionic behavior) dominate while in the region with negative $R$, attractive interactions (Bosonic behavior) dominate. Moreover, $\mid R \mid$ indicates the average size of fluctuations.

\section{Concluding Remarks}
\label{Sec5}
   In this paper, we extend our former research of charged topological Lovelock AdS black holes to the non-extended phase space. Specifically, we investigate phase transition of (n+1)-dimensional Lovelock AdS black holes in the grand canonical ensemble. Firstly, we calculated the specific heat at constant electric potential. To probe the impact of the various parameters, we utilize the control variate  method and solve the phase transition condition equation numerically for the case $k=1,-1$. There are two critical points for the case $n=6,k=1$ while there is only one for other cases. And the distance between the two phase transition points becomes larger with the increasing of $\Lambda$ and $\Phi$ while it first becomes larger then becomes smaller with the increasing of $\alpha$. We also study the behavior of specific heat graphically. As can be seen from the graph, the black holes can be divided into three phases. Namely, small stable ($C_\Phi>0$) black hole, medium unstable ($C_\Phi<0$) black hole and large stable ($C_\Phi>0$) black hole. The graph of Hawking temperature is also depicted to check whether the phase transition points locate in the physical region. For $k=0$, there exists no phase transition point.

   To figure out the nature of phase transition in the grand canonical ensemble, we carry out an analytic check of the analog form of Ehrenfest equations proposed by Banerjee et al. It is proved that the two Ehrenfest equations hold at the phase transition point. Prigogine-Defay ratio is also calculated. Based on these results, one can draw the conclusion that Lovelock AdS black holes in grand canonical ensemble undergo a second order phase transition.

   To examine the phase structure in the grand canonical ensemble, we also utilize the thermodynamic geometry method. Specifically, we calculate both the Weinhold metric and Ruppeiner metric. It is shown that in the denominator of Ruppeiner scalar curvature, the fifth factor is exactly one quarter of the denominator of the specific heat while the last factor coincides with the numerator of the Hawking temperature. So the Ruppeiner scalar curvature may diverge exactly where the specific heat diverges. It also reveals the extremal black hole condition that the Hawking temperature is zero. From the graph of Ruppeiner scalar curvature, one can see clearly that the divergence structures of both the Ruppeiner scalar curvature and the specific heat are exactly the same. Our research provides one more example that Ruppeiner metric serves as a wonderful tool to probe the phase structures of black holes.

   Note that one may vary the spatial dimension, the cosmological constant, and the coefficients of the curvature terms in the Lagrangian and we mainly concentrate on a few instances of a very large model in this paper. The control variate method has been utilized to crack down the problem of probing the impact of the various parameters. We choose such parameter regions that we can compare our results with those in former literatures. One can easily extend our results to more cases. The black hole solution here was derived for the special case that the second and third order Lovelock coefficients satisfy certain conditions. Phase transition in the non-extended space of more general black hole solutions in Lovelock gravtiy would be further investigated in our future work. Also note that the methods utilized in this paper can be generalized to an arbitrary non-linear electrodynamics Lagrangian, the specific results in this paper however are model-dependent. For a more general analysis, we would like to draw the readers' attention to the excellent work \cite{Miskovic99}, where the authors presented an elegant procedure for Gauss-Bonnet gravity regardless of the explicit form of the nonlinear electrodynamics Lagrangian. It certainly deserves to extend this treatment to the third-order Lovelock case in future research.

 \section*{Acknowledgements}
We would like to express our sincere gratitude to both the editor and the referee whose hard work have help improved the quality of this paper greatly. This research is supported by the National Natural Science
Foundation of China (Grant Nos.11235003, 11175019, 11178007). It is
also supported by \textquotedblleft Thousand Hundred
Ten\textquotedblright \,Project of Guangdong Province and supported by Department of Education of Guangdong Province (Grant No.2014KQNCX191).

\end{document}